\documentstyle[prl,aps,epsf,multicol,tabularx]{revtex}
\begin{document}

\title{Cluster mean-field study of the parity conserving phase transition}
\author{G\'eza \'Odor and Attila Szolnoki}
\address{Research Institute for Technical Physics and Materials Science, \\
H-1525 Budapest, P.O.Box 49, Hungary}    
\maketitle

\begin{abstract}
The phase transition of the even offspringed branching and
annihilating random walk is studied by $N$-cluster mean-field
approximations on one-dimensional lattices. By allowing to reach zero branching
rate a phase transition can be seen for any $N \le 12$.
The coherent anomaly extrapolations applied for the series of
approximations results in $\nu_{\perp}=1.85(3)$ and $\beta=0.96(2)$.

\end{abstract}
\begin{multicols}{2}

%----------------------------------------------------------------------------
\section{Introduction}
%----------------------------------------------------------------------------

The study of nonequilibrium phase transition universality classes is
one of the most fundamental tasks of statistical physics
\cite{DickMar,Hin2000,dok,Uweof,Racznewrev,svenrev}. 
Such phenomena may appear in models of population, epidemics, 
catalysis, cooperative transport \cite{DickMar} and enzyme biology 
\cite{Berry} for example. For a long time only the
universality class of the directed percolation (DP) \cite{DP} has been
known, but later it turned out that different classes may appear in
other models (for a review see \cite{dok}). The most prominent
example occurs in one-dimensional systems exhibiting $Z_2$ symmetrical 
absorbing states \cite{Gras84,Men94,meod95,BBE,Hin97} and parity 
conserving (PC) -- explicit or underlying -- branching and
annihilating random walk (BARWe) dynamics ($A\to 3A$, $2A\to\emptyset$)
of the dual variables ($A$) \cite{Taka,2-BAW,Park94,Cardy-Tauber}. 
For a review see \cite{MeOdof}. This class is also called as
directed Ising, DP2 or generalized voter model class.

However understanding these phenomena involving a satisfying solution
of the corresponding field theory is very rare.
For BARWe models field theory failed to give quantitatively precise
results in one dimension \cite{Cardy-Tauber} 
because systematic epsilon expansion
brakes down due to a second critical dimension at $d'_c=4/3$ below $d_c=2$.
Very recently an other field theory has been suggested and analyzed by
numerical simulation of the Langevin equation for systems exhibiting
$Z_2$ symmetric absorbing states \cite{HCDMlett}. 
Numerical approximations ranging from simulations
\cite{Hin97,2-BAW,MeOdof,Jensen,meod96} to series expansions \cite{JenGDK}, 
cluster mean-field approximations \cite{meod95,meod96} 
and empty interval method \cite{ZAM03} established the values of
critical exponents firmly.

Recently Zhong et al.\cite{ZAM03} constructed a special parity conserving
reaction-diffusion model (hereafter we call it ZAM model). 
By studying it up to pair approximations they claimed, that cluster 
mean-field studies fail to reproduce the phase transition of the BARWe
model if one considers clusters which are not large enough. 
Therefore one may arrive to the conclusion that cluster mean-field
and especially the site mean-field solution break down generally in case of
PC class transitions.
However in that study the branching attempt probability
($\sigma$) was fixed to a finite value, so one should not expect to see 
the mean-field transition, which is known to occur at zero
branching rate. 
In the present study we show that the cluster mean-field
approximations can describe the the mean-field
transition qualitatively well even for small cluster sizes if we
do not exclude the neighborhood of the zero branching rate from the 
parameter space.
These approximations are performed on one-dimensional clusters
therefore for large cluster sizes $N$ one expects to see a convergence 
towards the PC class transition at $\sigma_c>0$. We apply coherent
anomaly extrapolations for the sequence of cluster mean-field results
and give estimates for some exponents of the critical behavior
in one dimension.

The interplay of diffusion and fluctuation has already been shown in many
reaction-diffusion models
(see for example \cite{meod96,2340cikk,bin2dcikk,difflett}).
Although the diffusion is unable to change the universal behavior 
it can affect the location of the transition and even more it can
change the stability of a fixed point in case of competing reactions
\cite{2340cikk,bin2dcikk,difflett}).
 To investigate the possible role of diffusion we also extend the
 parameter space by modifying the diffusion rate.

%----------------------------------------------------------------------------
\section{The cluster mean-field method}
%----------------------------------------------------------------------------

The generalized (cluster) mean-field (GMF) method is an extension of the
usual mean-field calculation by setting up master equations
for $n$-point configuration probabilities of site variables 
$s_i\in\{A,\emptyset\}$
\begin{equation}
\frac{\partial P_n(\{s_i\})}{\partial t} = f\left (P_n(\{s_i\})\right) \ ,
\label{mastereq}
\end{equation}
where the function $f$ depends on the transition rules of
$\{P_m\}$ ``block probabilities'' at time $t$.

At the level of $N$-point approximation the correlations are
neglected for $n>N$, that is, $P_n(s_1,...,s_n)$ is expressed by using 
the Bayesian extension process \cite{gut87,dic88,SzSzB90},
\begin{equation}
P_{n}(s_1,...,s_n)={\prod_{j=0}^{j=n-N}
P_N(s_{1+j},...,s_{N+j}) \over \prod_{j=1}^{j=n-N} 
P_{N-1}(s_{1+j},...,s_{N-1+j})} \ .
\label{eq:bayes}
\end{equation}
In principle, $2^N-1$ parameters are required to define the
probability of all the $N$-point configurations. This
number, however, is drastically reduced by the following
conditions : In the stationary state the particle distribution is assumed 
to be symmetric with respect to translation and reflection. 
Furthermore, the block probability consistency results in:
\begin{eqnarray}
P_{n}(s_1,...,s_n) &=& \sum_{s_{n+1}} P_{n+1}
(s_1,...,s_n,s_{n+1}) \ , \nonumber \\
P_{n}(s_1,...,s_n) &=& \sum_{s_0} P_{n+1} (s_0,s_1,...,s_n)
\ . \label{eq:prob} \nonumber
\end{eqnarray}

Here we apply GMF for one-dimensional, site restricted 
lattice versions of BARWe. 
Taking into account spatial the symmetries in case of the $N=10$ 
GMF approximation one has to find the solution of equations of 528 independent 
variables. This has been achieved with the help of MATHEMATICA
software. We required 20 digit accuracy in the results and arbitrary
precision during the calculations.

It is well known that such approximations predict the phase structure 
qualitatively well in one dimension, 
provided $N$ is large enough to take into account
the relevant interaction terms.
For example $N>1$ is needed to take into account particle diffusion terms, 
while $N>2$ was found to be necessary in case of binary production processes 
involving pair induced reactions \cite{2340cikk,szolpcp}. The
GMF is an efficient phase diagram exploration method and although it is
set up for the $d=1$ lattice in previous cases it provided qualitatively 
good phase diagram for higher dimensional, mean-field versions as well 
(see example \cite{SzO94cikk,bin2dcikk,SzoCAM,SzoIsing}).

%----------------------------------------------------------------------------
\section{The ZAM model}
%----------------------------------------------------------------------------

In \cite{ZAM03} Zhong et al. defined a special, one-dimensional, parity 
conserving lattice model in which each site is either empty or singly 
occupied. 
The state of the lattice is updated asynchronously by the following rules:
an occupied site is chosen randomly and it is tried for diffusion, 
at rate $\Gamma$ (probability $\Gamma/(\Gamma+\Omega)$), 
or branching, at rate $\Omega$ (probability  $\Omega/(\Gamma+\Omega)$); 
while the time is increased by $1/N$, where $N$ is the number of occupied
(active) sites. In a diffusion step the particle jumps to its
  randomly chosen nearest neighbor sites. If the  
  site is occupied both particles are annihilated 
  with probability $r$. On the other hand the jump
  is rejected with probability $1-r$. The branching
  process involves the creation of two new particles
  around the neighborhood. If either, or both
  neighboring sites is previously occupied the target
  site(s) become empty with probability $r$. Otherwise
  the lattice remains unaltered with probability $1-r$.

\begin{figure}
\begin{center}
\epsfxsize=80mm
\centerline{\epsffile{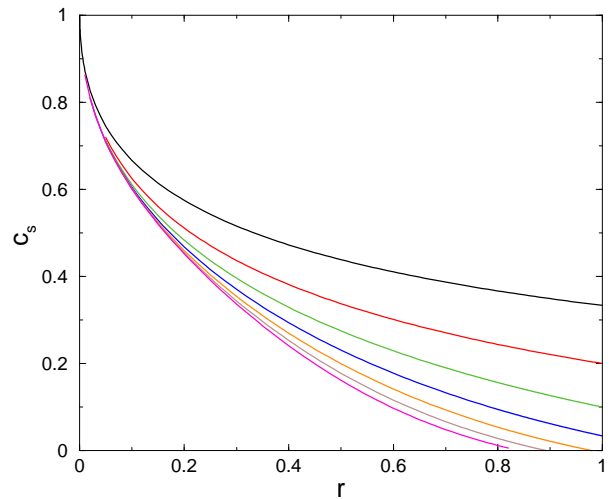}}
\caption{(Color online) GMF results for the steady state concentration 
in the ZAM model for $N=1,2,..7$ clusters (top to bottom).}
\label{fig1}
\end{center}
\end{figure}

The site mean-field equation for the concentration $c$ (probability of
site occupancy) is
\begin{equation}
\label{cdot1site}
\frac{d}{dt}c=-2rc^2+2c(1-c)^2-2rc^3\;,
\end{equation}
where $\Gamma=\Omega=1$ was taken, hence the branching probability is
fixed to $1/2$. Similarly the $N=2$ pair mean-field approximation was
solved and in both cases the system always evolves
to an active phase with finite concentration $c_s$ (see Fig.\ref{fig1}).
These results were compared with those of the $r<1$ rate model in one
dimension, for which PC class continuous phase transition is known at 
$r_c>0$ \cite{2-BAW}. The conclusion was drawn that cluster mean-field 
approximations of low orders fail to reproduce the phase diagram,
the convergence is very slow and the calculations are complicated.
However it is also known that in the mean-filed approximation -- which
is valid for $d>d_c=2$ dimensions -- the phase transition occurs at zero
branching rate \cite{Cardy-Tauber}, so it is not surprising that
by fixing $\Omega=1$ one does not see a phase transition in the
corresponding mean-field approximations. On the other hand Zhong et
al. found a phase transition for $0< r_c<1$ by the parity interval
method -- which also involves a mean-field like approximation -- 
set up for this one-dimensional model \cite{ZAM03}.
The authors acknowledged that it had been possible to see the phase
transition in related PC class models for larger ($N>2$) cluster
sizes \cite{meod95}. 

To clarify this we extended the cluster mean-field method for higher 
orders and for other versions of this BARWe model exhibiting 
PC class transition.
We determined the steady state solutions of Eq.(\ref{mastereq})
taking into account Eqs.(\ref{eq:bayes},\ref{eq:prob}) and calculated
the corresponding steady state densities $c_s(r)$ (see Fig.\ref{fig1}). 
Indeed higher levels of approximations ($N > 4$) of the ZAM model 
result in phase transition with $r_c(N)\le 1$ converging towards the
simulations ($r_c=0.470(5)$).

%----------------------------------------------------------------------------
\section{GMF results for the ZAMB model}
%----------------------------------------------------------------------------

As we saw in the preceding section cluster mean-field
approximations of the ZAM model give qualitatively good phase
diagram for $N>4$ and the steady state solutions converge to the
MC simulations.
However going much further with the GMF study of the ZAM model is
time consuming especially because the numerical root finding of
Eq.(\ref{mastereq}) gets computationally demanding for large number of 
variables.
On the other hand by modifying the ZAM model slightly in such a way
that zero branching rate is allowed with the restriction 
$ \sigma = 2 \Omega/(\Gamma+\Omega)= 1 - r$ (ZAMB) one immediately 
finds the expected mean-field transition at 
$\sigma_c = 0$ for $N\ge 1$.
\begin{figure}
\begin{center}
\epsfxsize=80mm
\centerline{\epsffile{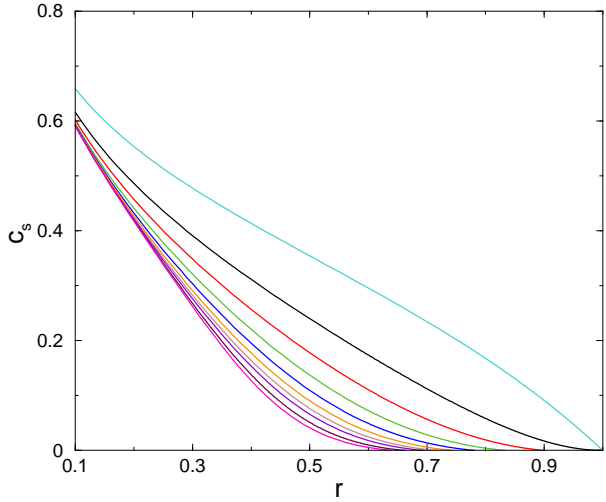}}
\caption{(Color online)Steady state concentration GMF results for the
ZAMB model for $N=1,2,3,4,5,6,7,8,10,12$ clusters (top to bottom).}
\label{fig2}
\end{center}
\end{figure}
The site approximation equation for the concentration is
\begin{equation}
\frac{d}{dt}c = 2\,c\,\left( 1 + {c^2}\,{{\left( r - 1 \right) }^2} - r - 
    c\,\left( 2 - \,r \right)  \right) \ ,
\label{zambeq} 
\end{equation}
exhibiting the steady state solution
\begin{equation}
c_s = {\frac{ 2 - r - {\sqrt{r}}\,{\sqrt{8 - 11\,r + 4\,{r^2}}} }
   {2\,{{\left( r - 1 \right) }^2}}} .
\end{equation}
for $r<1$. For $r=1$ Eq.(\ref{zambeq}) simplifies to
\begin{equation}
\frac{d}{dt}c = - 2 c^2
\end{equation}
resulting in $c\propto 1/t$ particle density decay in agreement with
the mean-field expectations. The supercritical behavior
in the active phase can be characterized by $\beta=1$ leading
order singularity for $r_c \le 1$ 
\begin{equation} 
c_s \propto |r-r_c|^{\beta} \label{truebeta}
\end{equation}
and $\beta'=2$ correction to  scaling exponent defined as
\begin{equation} 
c_s = a |r-r_c|^{\beta} + b |r-r_c|^{\beta'} \ .
\end{equation}

The steady-state solution has been determined for 
$N=1,2,3,4,5,6,7,8,10,12$ (see Fig.\ref{fig2}). As one can see
even the pair approximation gives $r_c(2) = 1$, but for $N>2$ the
transition point starts shifting towards the true transition point
$r_c < 1$ as expected in one dimension.
One can also observe a concave shape of the curves corresponding to
the corrections to scaling with $\beta'=2$ exponent.

\begin{figure}
\begin{center}
\epsfxsize=80mm
\centerline{\epsffile{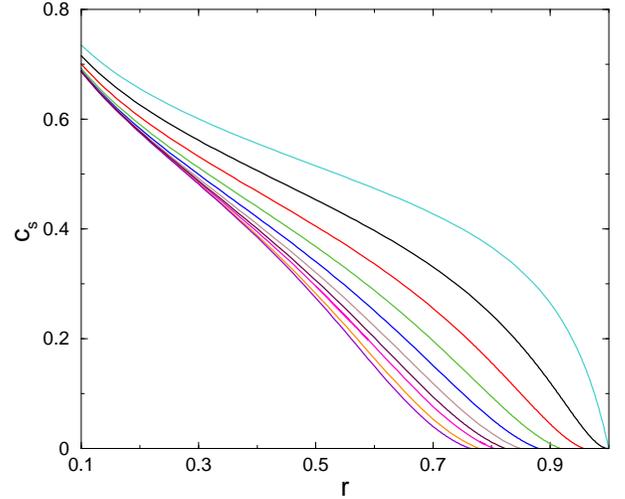}}
\caption{(Color online)Steady state concentration GMF results for the
ZAMB model for $D=0.2$ diffusion and $N=1,2,3,4,5,6,7,8,10,12$ 
clusters (top to bottom)}
\label{fig3}
\end{center}
\end{figure}

%----------------------------------------------------------------------------
\subsection{The effect of diffusion}
%----------------------------------------------------------------------------

In previous papers 
\cite{meod95,OBScikk,SzO94cikk,2340cikk,bin2dcikk,difflett}) it was shown that
the diffusion strength can cause relevant effects on the phase diagram
of reaction-diffusion models when it competes with the reactions.
This is reflected in the cluster mean-field approximations in such a
way that stronger diffusion ``washes out'' fluctuations and causes a
transition, which is more site mean-field like, 
while higher $N$ takes into account more fluctuations, 
hence opposes the effects of the diffusion. 
Here we investigated the effect of diffusion by lowering the hopping 
probability of the ZAMB model to $D = 2 \Gamma/(\Gamma + \Omega) = 0.2$. 
The steady state results (Fig.\ref{fig3}) show that the concentration 
curves arrive with higher slopes to the $c_s=0$ axis than in case of $D=1$. 
This permits us to obtain $a(N)$ more precisely since the relative error
of the amplitudes is smaller and the quadratic correction to scaling
is weaker. As the consequence the numerical root finding is not so
much affected by basin of attraction of the absorbing state fixed
point solution.
Note that in the pair approximation the linear amplitudes $a(2)$ are zero.

%----------------------------------------------------------------------------
\section{Simulation results}
%----------------------------------------------------------------------------

By applying fitting with the expected scaling form
\begin{equation}
|r_c(N)-r_c|^{\nu_{\perp}} \propto 1/N \label{rcscale}
\end{equation}
to the GMF data one can determine the location of the transition and 
$\nu_{\perp}$ simultaneously. For $D=1$ this gives $r_c=0.402$, while 
for $D=0,2$  $r_c=0.65$. 
However to obtain better critical value estimates we used more precise 
$r_c$ values in the fitting procedure, which can be 
deduced from simulations.
\begin{figure}
\begin{center}
\epsfxsize=80mm
\centerline{\epsffile{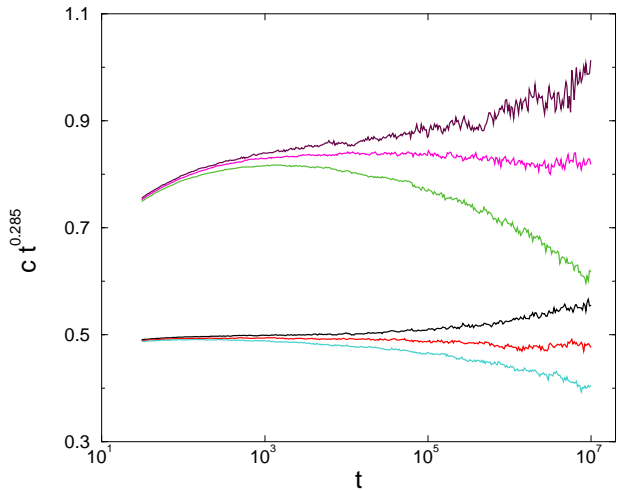}}
\caption{(Color online) Particle density decay of the ZAMB model times
the expected critical power-law ($t^{0.285}$). Different curves
correspond to $r=0.565, 0.562, 0.56$ and $D=0.2$,
$r=0.408, 0.409, 0.41$ and $D=1$ (from top to bottom).}
\label{figMC}
\end{center}
\end{figure}
The simulations were performed on one-dimensional lattices of sizes $L=10^7$
with periodic boundary conditions. The runs were started from half filled
lattices with randomly distributed particles. One elementary Monte
Carlo step (MCS) consists of the following processes.
A particle and a direction are selected randomly. If the nearest
neighbour (nn) in the selected direction was empty the particle moves
to it (with probability $D$). If it was filled, both particles
are removed (with probability $r$). The time -- measured by MCS ---
is updated by $1/n_P$,
where $n_P$ is the total particle number at time $t$. To perform the 
branching another particle is selected randomly (with probability $1-r$).
Depending on the status of the two nn ($s_{i-1}$,$s_{i+1}$) the
following process may occur.

\noindent a) If $s_{i-1} = s_{i+1} = \emptyset$, two new particles
are created.

\noindent b) If $s_{i-1} = s_{i+1} = A$ the two nn particles are
removed with probability $r$.

\noindent c) If $s_{i-1} \ne s_{i+1} $ the nn are swapped 
with probability $r$.
The time ($t$) is updated by $1/n_P$ again. 
The simulations were followed up to $t=10^7$ (MCS) or until 
$n_P=0$ (absorbing state). The concentration of particles $c_s(t)$ 
times the expected density decay power-law  $1/c\propto t^{0.285}$ 
\cite{MeOdof,Jensen} is plotted  on Fig.\ref{figMC}.
One can read-off $r_c=0.409(1)$ for $D=1$ and $r_c=0.562(1)$ for $D=0.2$.
For $D=1$ this agrees well with the extrapolation
results of (Eq.(\ref{rcscale})), but for $D=0.2$ the deviation is not
negligible.

%----------------------------------------------------------------------------
\section{Coherent anomaly extrapolations}
%---------------------------------------------------------------------------- 

\begin{figure}
\begin{center}
\epsfxsize=80mm
\centerline{\epsffile{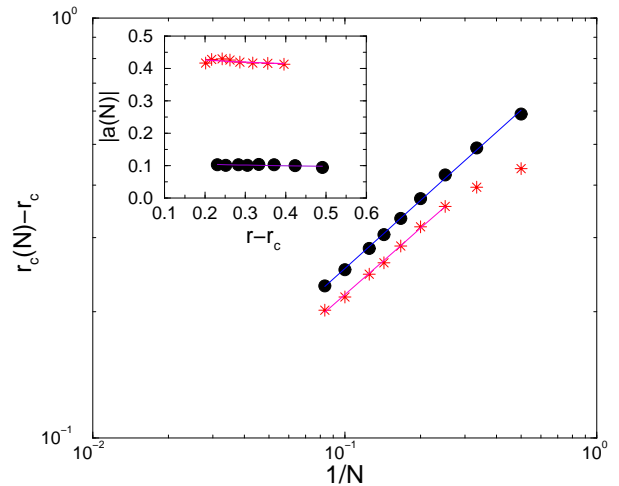}}
\caption{(Color online) CAM extrapolation results for the critical
point $r_c(N)$ ($2 \le N \le 12$). 
The insert shows $|a(N)|$ ($3 \le N \le 12$).
Stars correspond to $D=1$, boxes to $D=0.2$. Numerical errors are
smaller than the symbol sizes.}
\label{figcam}
\end{center}
\end{figure}

According to scaling theory the location of the critical point for sizes
$N$ ($r_c(N)$) scales (in the large $N$ limit) as Eq.(\ref{rcscale}).
Precise extrapolation can be obtained by applying the critical
transition point values of simulation in the mentioned scaling form.
The $r_c(N)$ for the $N$-th level of approximations 
was determined by quadratic fitting for $c_s(N) < 0.002$. 
Figure \ref{figcam} shows $r_c(N)$ as the function of $1/N$.
The fit of form (\ref{rcscale}) yields $1/\nu_{\perp} =0.54(1)$  
for $D=1$ and  $1/\nu_{\perp} = 0.53(1)$ for $D=0.2$ 
(see Fig.\ref{figcam}). The value $\nu_{\perp}=1.85(3)$, agrees well with the
value from the literature $\nu_{\perp}=1.84(6)$ \cite{Jensen} for this class.

We applied the coherent anomaly method (CAM) \cite{suz86} for the
$N$-cluster GMF results to extrapolate to the $N\to\infty$ behavior.
This method has been proven to be successful for obtaining critical 
exponents of nonequilibrium absorbing phase transitions 
\cite{meod95,meod96,OGDP,pcpdcikk,coagcikk,SzoCAM,ParkCAM}.
Earlier the GMF+CAM method was already used for analyzing the PC class 
transition of the NEKIM \cite{meod95,meod96} up to $N\le 6$ cluster
sizes. That study arrived to the rough estimate $\beta \simeq 1$, 
which agrees with the the series expansion result $\beta=1.00(5)$ 
\cite{Jensen97} and the simulation results marginally 
$\beta=0.95(2)$ \cite{MeOdof}.
Now we apply this method for cluster sizes up to $N\le 12$ and improve 
the GMF+CAM results \cite{meod95,meod96} for a model PC exhibiting PC
class transition.
According to CAM the amplitudes $a(N)$ of the cluster mean-field 
singularities
\begin{equation}
c_s(N) = |a(N)| | r_c(N) - r |^{\beta_{MF}} \label{gmfscal}
\end{equation}
scale in such a way 
\begin{equation}
|a(N)| \propto |r_c(N)-r_c|^{\beta - \beta_{MF}} \label{anoscal}
\end{equation}
that the exponent of the true singular behavior (Eq.(\ref{truebeta}))
can be estimated. 
The $a(N)$ amplitudes were determined by
linear fitting to the local slopes of the $c_s(N)$ data in the 
neighborhood of $r_c(N)$.
The amplitudes are shown in Table \ref{tab} and in the insert of 
Fig. \ref{figcam}.
The fitting using form Eq.(\ref{anoscal}) for $N>2$ data (since
$a(2)=0$) results in $\beta=0.92(5)$ for $D=1$ and  
$\beta=0.96(2)$ for $D=0.2$.
The CAM for $D=1$ results in bigger numerical error than for $D=0.2$, 
because for $D=1$ the quadratic corrections to scaling are stronger
and the $c_s(r)$ numerical solutions are affected by the attractive
basin of the neighboring absorbing state fixed point.
These values agree well with the simulation results 
$\beta=0.95(2)$ \cite{MeOdof}, $\beta=0.94(6)$ \cite{Jensen} as well
as with that of the parity interval method $\beta=0.92(2)$ \cite{ZAM03}.

%----------------------------------------------------------------------------
\section{Conclusions}
%----------------------------------------------------------------------------

We showed that even low order GMF approximations can describe the phase 
transition of an even-offspringed BARW model correctly if one allows 
appropriate parameterization. We applied GMF approximations for a one 
dimensional site restricted lattice model. 
By allowing to reach the zero branching rate 
in the ZAM model we showed that the site mean-field solution is in
agreement with that of the field theory for this class
\cite{Cardy-Tauber}.
Note that this kind of analysis resulted in similar steady-state
solutions in case of an other PC class model \cite{meod95,meod96},
although there the branching rate can't be read-off explicitly.

The GMF approximations were determined up to $N=12$ and convergence
towards the simulation results were shown. Using scaling and CAM
theory we obtained $\nu_{\perp}=1.85(3)$ and $\beta=0.96(2)$ critical
value estimates matching the best precision available in the literature
for the PC universality class.

{\bf Acknowledgements:}

The authors thank the access to the NIIFI Cluster-GRID, LCG-GRID 
and to the Supercomputer Center of Hungary.
Support from the Hungarian research fund OTKA (Grant Nos. T-046129,
T47003) is acknowledged.

\begin{center}
\begin{table}[h]
\begin{tabular}{lllll} 
\hline
\hline
$n \backslash D$ \,& \hspace{1.0cm} $0.2$ & & \hspace{1.2cm} & $1$ \\
  & $r_c(N)$ & $|a(N)|$ & $r_c(N)$ & $|a(N)|$ \\
\hline
1  & 1 & 5 &  1 & 1  \\
2  & 1 & 0 &  1 & 0 \\
3  & 0.9575(2) & 0.4124(4) &  0.8997(2) & 0.094(5) \\
4  & 0.9177(1) & 0.4155(3) &  0.8321(1) & 0.100(8) \\
5  & 0.8802(4) & 0.4167(9) &  0.7805(1) & 0.103(7) \\
6  & 0.8485(1) & 0.4192(9) &  0.7423(9) & 0.104(7) \\
7  & 0.8235(1) & 0.4254(3) &  0.714(7) & 0.100(12) \\
8  & 0.8044(1) & 0.429(3) & 0.691(13) & 0.103(9) \\
10 & 0.7787(1) & 0.427(2) & 0.6608(1) & 0.100(6) \\
12 & 0.7633(2) & 0.416(2) & 0.6394(1) &  0.103(6) \\
\hline
$r_c$ & 0.562(1) &  &\,\,\, 0.409(1) & \\
$\nu_{\perp}$  & 1.88(3) &  &  & 1.85(3) \\
$\beta$ &  & 0.96(2) &  & 0.92(5) \\
\hline
\hline
\end{tabular}
\vspace{2mm}
\caption{\sf Summary of results for the ZAMB model}
\label{tab}
\end{table}
\end{center}

\end{multicols}
\end{document}